\documentclass[letterpaper, 10 pt, conference]{ieeeconf}  

\IEEEoverridecommandlockouts                              
\usepackage{hyperref}
\usepackage{amsmath}
\usepackage{amssymb}
\usepackage{amsfonts}
\usepackage{commath}
\usepackage{cite}
\usepackage{balance}
\usepackage{graphicx}
\usepackage{xcolor}
\usepackage{siunitx}
\usepackage{hyperref}
\usepackage{float}
\usepackage{algorithm}
\usepackage{algpseudocode}  
\usepackage{booktabs}
\usepackage{courier}
\newtheorem{theorem}{\textbf{Theorem}}

\newtheorem{remark}{{Remark}}
\newtheorem{assumption}{\textbf{Assumption}}
\newtheorem{definition}{\textbf{Definition}}
\newtheorem{problem}{\textbf{Problem}}

\title{\LARGE \bf
Set-Theoretic Direct Data-driven Predictive Control
}

\author{Mohammad Bajelani, Walter Lucia, and Klaske van Heusden
\thanks{We acknowledge the support of the Natural Sciences and Engineering Research Council of Canada (NSERC) [RGPIN-2023-03660].}
\thanks{Mohammad Bajelani and Klaske van Heusden are with the University of British Columbia, School of Engineering, 3333 University Way, Kelowna, BC V1V 1V7, Canada} 
\thanks{Walter Lucia is with the Concordia Institute for Information Systems Engineering (CIISE), Concordia University, Montreal, QC, H3G 1M8, Canada 
{\tt\small mohammad.bajelani, klaske.vanheusden@ubc.ca, walter.lucia@concordia.ca}}}

\begin{document}

\maketitle
\thispagestyle{empty}
\pagestyle{empty}

\begin{abstract}
    Designing the terminal ingredients of direct data-driven predictive control presents challenges due to its reliance on an implicit, non-minimal input-output data-driven representation. By considering the class of constrained LTI systems with unknown time delays, we propose a set-theoretic direct data-driven  predictive controller that does not require a terminal cost to provide closed-loop guarantees. In particular, first, starting from input/output data series, we propose a sample-based method to build N-step input output backward reachable sets. Then, we leverage the constructed family of backward reachable sets to derive a data-driven control law. The proposed method guarantees finite-time convergence and recursive feasibility, independent of objective function tuning. It requires neither explicit state estimation nor an explicit prediction model, relying solely on input-output measurements; therefore, unmodeled dynamics can be avoided. Finally, a numerical example highlights the effectiveness of the proposed method in stabilizing the system, whereas direct data-driven predictive control without terminal ingredients fails under the same conditions.
\end{abstract}

\section{Introduction}

Terminal costs and constraints, often termed terminal ingredients, have been proposed in Model Predictive Controllers (MPC) to approximate the gap between finite-time and infinite-time predictions, thereby ensuring closed-loop guarantees such as stability and recursive feasibility. In the absence of terminal ingredients, the stability of Data-Driven Predictive Control (DDPC), similar to MPC, depends on the prediction horizon and tuning of the objective function, see \cite[Ch.\ 12, E.g.\ 12.2]{borrelli2017predictive}. On the other hand, introducing conservative choice of terminal ingredients, such as the equilibrium point, may significantly decrease the Region of Attraction (RoA). Designing such ingredients for DDPC differs from MPC because only input-output measurements and a Hankel-based matrix representation are available. Current solutions for the terminal constraint are limited to the equilibrium point of the system \cite{berberich2021data}, artificial set points \cite{berberich2020data}, or require the system's lag to be known \cite{berberich2021design}. For a comprehensive discussion on this issue, refer to \cite[Sec 3.1]{berberich2024overview}, which highlights that designing the terminal ingredients in this framework is still an open question. In response to these limitations, we leverage reachability analysis as a systematic approach for designing terminal ingredients.

A prevalent approach for designing and analyzing constraint control systems is reachability analysis, which systematically explores all potential solutions to prevent constraint violations. Forward Reachable Sets (FRS) and Backward Reachable Sets (BRS) serve as fundamental tools in this context, enabling the calculation of states that can be reached from specified initial conditions or directed towards a target set over finite (or infinite) time. Several methodologies currently exist for computing these sets, including sampling-based methods, simulation-based techniques, set propagation, and Hamilton-Jacobi analysis, as detailed in \cite{lew2021sampling, donze2007systematic, bansal2017hamilton, althoff2021set}. Each methodology presents trade-offs between computational complexity and the accuracy of the resulting set representations. 

Most reachability analysis techniques rely on pre-specified state-space models, necessitating state estimation or full state measurement, which can pose challenges for practical implementation when often output measurements are available. Requiring \emph{input-state} data and \emph{exact knowledge of the system's order}, recursive matrix zonotopes are proposed in \cite{alanwar2023data} to calculate data-driven forward reachable sets. This approach allows for the utilization of a set of models instead of relying on a potentially inaccurate single model, effectively capturing the true dynamics of the system. Similarly, matrix zonotopes are employed in \cite{attar2023data} to calculate data-driven backward reachable sets. Using the data-driven over-approximated FRS and under-approximated BRS, two predictive controllers are introduced in \cite{alanwar2022robust} and \cite{attar2023data}. Note that any model assumptions made during the identification process, including the \emph{system's order}, may result in model mismatch, even in LTI systems \cite{martin2023guarantees}. In contrast, the input-output data-driven framework introduced by J. C. Willems in \cite{willems2005note} remains unaffected by unmodeled dynamics and unstructured uncertainty, as it defines the system using input-output data without relying on explicit representations.

Limited research has focused on reachability analysis when only input-output data is available and the number of states is unknown. In \cite{zhang2024data}, input-output safe sets are computed for iterative tasks, which require optimizing performance for a specific initial condition and objective function. In \cite{bajelani2024raw}, two online and offline methods are proposed to safely expand the input-output safe set of a short-sighted safety filter. Additionally, \cite{ossareh2023data} calculates maximal admissible sets for a constant input to design a data-driven reference governor. To the best of our knowledge, no method presently exists for calculating N-step Input-Output Backward Reachable Sets (N-IOBRS) from an implicit input-output data-driven representation, nor for employing these sets within a direct data-driven predictive control framework.

\begin{figure}[t]
    \centering
    \includegraphics[trim=1cm 0cm 0cm 1cm, 
    clip,width=0.75\linewidth]{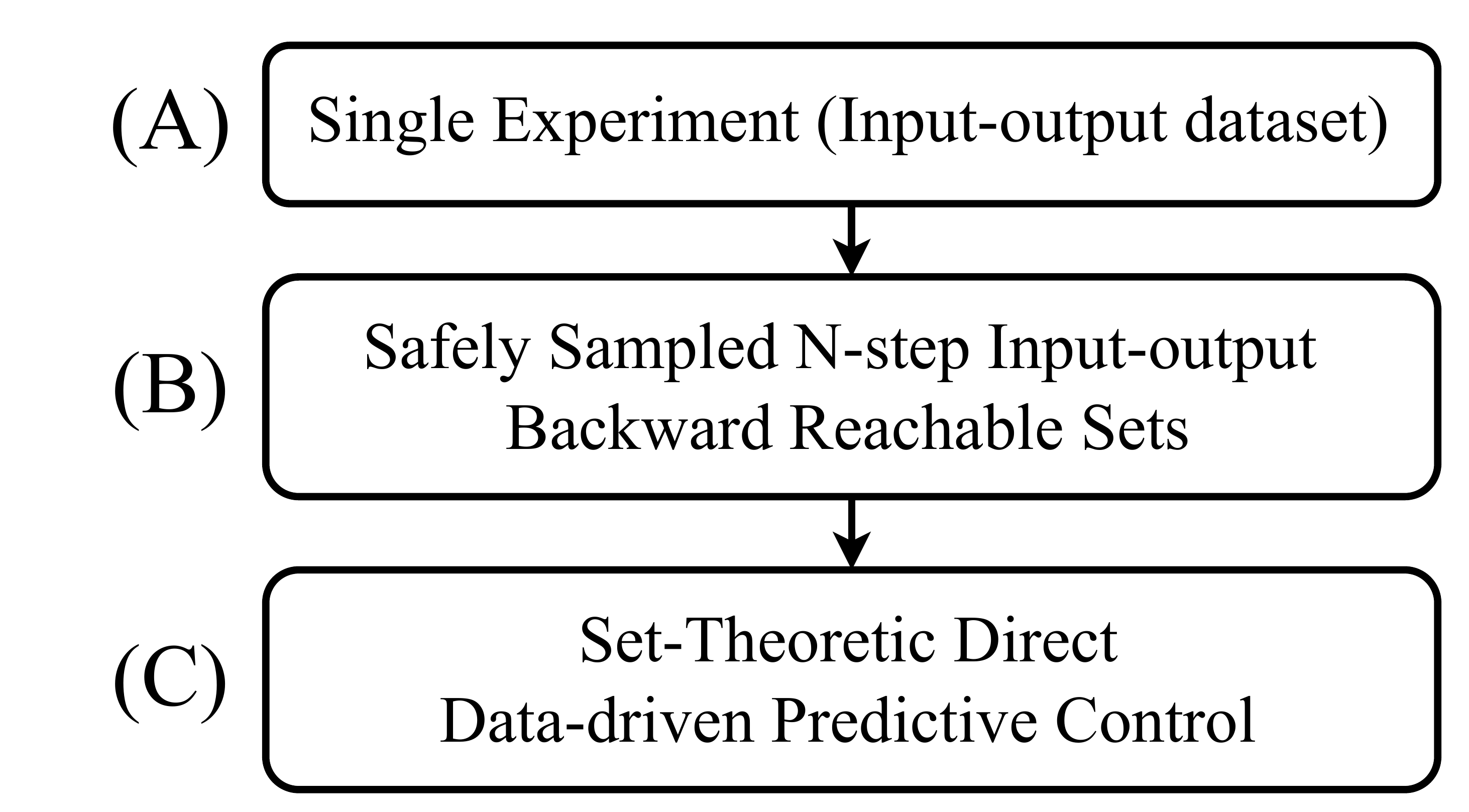}
    \caption{Flowchart of the overall process: (A) a single experiment generating an input-output dataset; (B) the sample-based method for computing N-IOBRS; and (C) the ST-DDPC developed from the sample-based N-IOBRS.}
    \label{fig: visual abstract}
    \vspace{-0.5cm}
\end{figure}

In this paper, first, we estimate N-IOBRS using a data-driven safety filter. Next, we utilize the resulting family of nested N-IOBRS to develop a direct data-driven predictive controller. The proposed method extends the state-feedback set-theoretic controller \cite{attar2023data} into the input-output framework while maintaining finite-time convergence and recursive feasibility properties. In contrast to \cite{attar2023data}, our method employs input-output multi-step prediction and attains a large region of attraction without requiring the exact system order, full state measurements, and an explicit system representation. It is important to emphasize that the proposed method systematically addresses input delays by over-approximating the system's lag, as described in \cite{bajelani2024raw}. To extend the proposed method to noisy measurements and input disturbances settings, the N-IOBRS must be appropriately tightened based on the noise and disturbance levels, which is beyond the scope of this paper. Figure \ref{fig: visual abstract} illustrates a flowchart depicting the overall process of the proposed method.

The remainder of the paper is organized as follows: Section \ref{sec: PRELIMINARIES} explains the preliminary materials and the problem statement. Section \ref{sec: N-IOBRS} presents the sample-based method to compute N-IOBRS. Section \ref{sec: ST-DDPC} provides the proposed controller, along with a proof of recursive feasibility and convergence. Numerical results and discussion are presented in Section \ref{sec: results}. Finally, Section \ref{sec: Conclusion} provides concluding remarks.

\section{Preliminaries and Problem Statement} \label{sec: PRELIMINARIES}

This section provides an overview of the preliminary materials related to set-theoretic predictive controllers and the input-output data-driven framework, as well as the assumptions regarding the underlying system and the specific problem of interest

\subsection{Constrained LTI systems}

Assume the underlying system is a deterministic discrete-time LTI system, defined in minimal state-space form with polytopic constraints, as follows:
\begin{equation} \label{eq:LTI-system}
    x_{t+1} = A x_t + B u_t, \quad y_t = C x_t + D u_t,
\end{equation}
\begin{equation} \label{eq:polytopic-constraints}
    u_t \in \mathcal{U}, \quad x_t \in \mathcal{X}, \quad y_t \in \mathcal{Y},
\end{equation}
where $x_t \in \mathbb{R}^n$, $u_t \in \mathbb{R}^m$, and $y_t \in \mathbb{R}^p$ are the state, input, and output vectors at time step $t$, respectively. The sets $\mathcal{U}$, $\mathcal{X}$, and $\mathcal{Y}$ are polytopes that define the admissible input, state, and output constraints. Assume the pair $(A,B)$ is controllable, and the pair $(A,C)$ is observable. Note that the tuple \((A,B,C,D)\) is unknown to both its values and dimensions. The only prior knowledge of the system consists of an informative single \emph{input-output} trajectory from the system \eqref{eq:LTI-system} and an upper bound on the system's lag.

\begin{definition}[System's Lag \cite{9654975}] \label{definition: lag} 
    $\underline{l}=l(A, C)$ denotes the lag of the system (\ref{eq:LTI-system}), in which $l(A, C)$ is the smallest integer that can make the observability matrix full rank.
    \begin{equation*}
        l(A, C):=(C, C A, \ldots, C A^{l-1}).
    \end{equation*}
\end{definition}

\begin{definition}[LTI System's Trajectory \cite{9654975}] \label{Def: LTI system's trajectory}
    Let \( G \) be an LTI system with minimal realization \((A,B,C,D)\). The sequence \(\{u_t,y_t\}_{t=0}^{N-1}\) is an input-output sequence of this system if there exists an initial condition \( x_{0} \in \mathbb{R}^n \) and a state sequence \(\{x_t\}_{t=0}^{N}\) such that 
    \begin{equation*}
    \begin{aligned}
        &x_{t+1} = A x_{t} + B u_{t}, \\
        &y_{t} = C x_{t} + D u_{t}, \quad \forall t \in \{0, 1, 2, \ldots, N-1\}.
    \end{aligned}
    \end{equation*}     
\end{definition}

\begin{definition}[Convex Hull] \label{def:convexhull}
The convex hull of set of $k$ points in $\mathcal{S} \subseteq \mathbb{R}^n$ is defined as:
\begin{equation*} \label{eq:convexhull}
    \text{conv}(\mathcal{S}) := \{ \sum_{i=1}^{k} \lambda_i v_i \mid v_i \in \mathcal{S}, \lambda_i \geq 0, \sum_{i=1}^{k} \lambda_i = 1 \}.
\end{equation*}
\end{definition}

\begin{definition}[Polytope]  \label{def:polytope-HV}
    A polytope, $\mathcal{P}$, can be defined by half spaces ($H$-representation $H(\mathcal{P})$) or its vertices ($V$-representation $V(\mathcal{P})$) as follows, respectively:
    \begin{equation*} \label{eq:polytope-H}
        H(\mathcal{P}) := \left\{ x \in \mathbb{R}^n \mid C x \leq d, \; C \in \mathbb{R}^{q \times n}, \; d \in \mathbb{R}^{q} \right\},
    \end{equation*}
    \begin{equation*} \label{eq:polytope-V}
        V(\mathcal{P}) := \text{conv}\left( \{v_1, v_2, \dots, v_k \}\right), \quad v_i \in \mathbb{R}^n.
    \end{equation*}
\end{definition}

\subsection{Input-output framework}

Assume that an input-output trajectory of the system \eqref{eq:LTI-system}, as defined in Definition \ref{Def: LTI system's trajectory}, with length \(N_0\) is available in the form of the following vectors:
\begin{subequations}
    \begin{equation}
    u_{\left[0, N_0-1\right]}^d = \left[u_0^{\top}, \ldots, u_{N_0-1}^{\top}\right]^{\top} \in \mathbb{R}^{mN_0\times1},
    \end{equation}
    \begin{equation}
    y_{\left[0, N_0-1\right]}^d = \left[y_0^{\top}, \ldots, y_{N_0-1}^{\top}\right]^{\top} \in  \mathbb{R}^{pN_0\times1}.
    \end{equation}
\end{subequations}

The Hankel matrices \(H_L(u^d) \in \mathbb{R}^{(mL) \times (N_0 - L + 1)}\) and \(H_L(y^d) \in \mathbb{R}^{(pL) \times (N_0 - L + 1)}\), corresponding to the given input-output trajectory and consisting of \(mL\) and \(pL\) rows, are defined as follows:
\begin{subequations}
    \begin{equation}
    H_L(u^d) = \left[\begin{array}{cccc}
    u_0 & u_1 & \cdots & u_{N_0-L} \\
    u_1 & u_2 & \cdots & u_{N_0-L+1} \\
    \vdots & \vdots & \ddots & \vdots \\
    u_{L-1} & u_L & \cdots & u_{N_0-1}
    \end{array}\right] ,
    \end{equation}
    \begin{equation}
    H_L(y^d) = \left[\begin{array}{cccc}
    y_0 & y_1 & \cdots & y_{N_0-L} \\
    y_1 & y_2 & \cdots & y_{N_0-L+1} \\
    \vdots & \vdots & \ddots & \vdots \\
    y_{L-1} & y_L & \cdots & y_{N_0-1}
    \end{array}\right].
    \end{equation}
\end{subequations}

\begin{definition}[Persistently Excitation] \label{Def: PE condition}
    Let the Hankel matrix's rank be $rank(H_L(u)) = mL$, then $u \in \mathbb{R}^m$ represents a persistently exciting signal of order $L$.
\end{definition}

\begin{theorem}[Fundamental Lemma \cite{berberich2020robust}] \label{Fundamental Lemma}
    Let $u^d$ be persistently exciting of order $L+n$, and ${\{u^d_t},{y^d_t\}}_{t=0}^{N_{0}-1}$ a trajectory of system $G$. Then, ${\{\bar{u}},{\bar{y}}\}$ is a trajectory of system $G$ if and only if there exists $\alpha \in \mathbb{R}^{N_{0}-L+1}$ such that
    \begin{equation} \label{BST model}
        \left[\begin{array}{l}
        H_L\left(u^d\right) \\
        H_L\left(y^d\right)
        \end{array}\right] \alpha=\left[\begin{array}{l}
        \bar{u} \\
        \bar{y}
        \end{array}\right].  \vspace{0.1cm} 
    \end{equation}
\end{theorem}

The fundamental lemma indicates that all input-output trajectories of a discrete-time LTI system can be spanned by Hankel matrix columns. This enables us to directly design a control law in the input-output framework using raw data \cite{yang2015data, 8795639}, eliminating the requirement to find the underlying system \eqref{eq:LTI-system}. Irrespective of the format in which the pre-recorded dataset is presented—whether as a Hankel matrix \cite{willems2005note}, a page matrix \cite{coulson2021distributionally}, or a collection of experimental results \cite{9654975}—the trajectory space can be fully spanned by these sequences, provided that the persistent excitation condition is satisfied. This condition enables the use of one or more experiments, from which random segments can be selected to represent the system's dynamics through input-output measurments.

To implicitly establish the initial condition and predict the system's behavior, the Hankel matrices must be divided into two sections. The first \(mT_{\text{ini}}\) rows in \(H_L(u^d)\) and \(pT_{\text{ini}}\) rows in \(H_L(y^d)\) correspond to past measurements (also known as past data), which is used to determine the initial condition, while the remaining rows, utilized to predict the system's output (also known as future data). For any choice of \(T_{\text{ini}} \geq \underline{l}\), the initial condition and system's order are implicitly determined. In other words, considering a sufficiently long segment of the past input-output trajectory can express the history of the dynamics of system \eqref{eq:LTI-system}, thereby eliminating the need for the underlying state, as noted in \cite[lemma 1]{markovsky2008data}.

\begin{figure}[b]
    \centering
    \includegraphics[trim=0.3cm 0cm 0.3cm 0cm, 
    clip,width=1\linewidth]{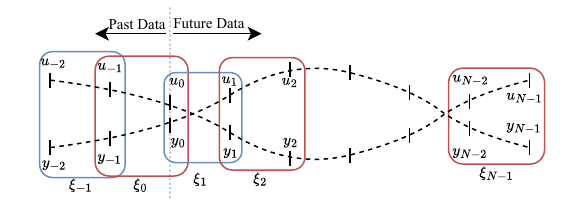}
    \caption{Visualization of an extended trajectory for an input-output trajectory, assuming $T_{\text{ini}}=2$, integrating both past data (historical measurements) and future data (predicted trajectory).}
    \label{fig: Extended Trajectory}
\end{figure}

We generalize the notion of the past input-output trajectory, referred to as the extended state in Definition \ref{definition: Extended State}, to any input-output trajectory through the concept of the extended trajectory, as defined in Definition \ref{definition: Extended Trajectory}. This enables us to define the extended state over predicted input-output trajectories, thus establishing a link to reachability analysis within the input-output framework and removing the need for the underlying state. Fig. \ref{fig: Extended Trajectory} visualizes how an extended trajectory is defined for an input-output trajectory consisting of past and future data.

\begin{definition}[Extended State\cite{berberich2021design}] \label{definition: Extended State}
    For some integers $T_{\text{ini}}\geq\underline{l}~$, the extended state $\xi$ at time $t$ is defined as follows
    \begin{align}\label{eq:extended_state}
    \xi(t) := \begin{bmatrix}u_{[t-T_{\text{ini}},t-1]}\\y_{[t-T_{\text{ini}},t-1]}\end{bmatrix} \in \mathbb{R}^{(m+p)T_{\text{ini}} \times 1} .
    \end{align}
    where $u_{[t-T_{\text{ini}},t-1]}$ and $y_{[t-T_{\text{ini}},t-1]}$ denote the last $T_{\text{ini}}$ input and output measurements at time $t$.
\end{definition}

\begin{definition}[Extended Trajectory] \label{definition: Extended Trajectory}
    For an input-output trajectory, $\left\{{u}_t, {y}_t\right\}_{t=-T_{\text{ini}}}^{N-1}$, the extended trajectory is defined as follows:\vspace{-0.3cm}
    \begin{equation*}\label{eq: Extended Trajectory}
        \bar{\xi} := [\xi_{-T_{\text{ini}}}, \ldots, \xi_{N-2}], \text{ where } \xi_t := \begin{bmatrix}u_{[t,t+T_{\text{ini}}-1]}\\y_{[t,t+T_{\text{ini}}-1]}\end{bmatrix}. \vspace{0.1cm}
    \end{equation*}
\end{definition}

\begin{definition}[Input-output Control Invariant Set] \label{def: Input-Output Control Invariant Set}
    A set \(\Xi \subseteq \mathbb{R}^{(m+p)T_{\text{ini}}}\) is defined as a control invariant set for the system (\ref{eq:LTI-system}) from input-output perspective if \vspace{-0.2cm}
    \begin{equation*}
        \xi(t_0) \in \Xi \Rightarrow \exists u(t_0) \in \mathcal{U} \text{ such that } \xi(t_0+1) \in \Xi, \forall t \geq t_0.    
    \end{equation*}
\end{definition}

\begin{definition} [Input-output Equilibrium Point] \label{def: Equilibrium point.}
    The extended state $\xi^s \in \mathbb{R}^{(m+p)T_{\text{ini}}}$ is an input-output equilibrium point of system (\ref{eq:LTI-system}) if it is defined by the sequence $\left\{{u}_t, {y}_t\right\}_{t=0}^{t=T_{\text{ini}}-1}$ with a constant value $\left({u}_t, {y}_t\right)=\left(u^s, y^s\right)$.
\end{definition}

The Definition \ref{def: Input-Output Control Invariant Set} implies that if the extended state can stay forever within a set in $\mathbb{R}^{(m+p)T_{\text{ini}}}$, then the underlying state $x$ stays forever within a set in $\mathbb{R}^n$. This relationship holds because, for any past input-output trajectory of the system \eqref{eq:LTI-system}, a unique $x \in \mathbb{R}^n$ can be determined if the matrices $(A, B, C, D)$ are known, as detailed in \cite[lemma 1]{markovsky2008data}.
\begin{definition} [N-IOBRS] \label{def: N-IOBRS}
    For a given target set of extended states, $\mathcal{T} \subseteq \mathbb{R}^{(m+p)T_{\text{ini}}}$, we define the $N$-step input-output backward reachable set as a set of extended states, named as $\Xi_N^\mathcal{T} \subseteq \mathbb{R}^{(m+p)T_{\text{ini}}}$, if for each $\xi_0 \in \Xi_N^\mathcal{T}$ there exists a sequence of inputs $\left\{{u}_t\right\}_{t=0}^{N-1}$ such that for some $t \in \{0,..., N-1\}$, $\xi_t \in \mathcal{T}$. More formally, \vspace{-0.2cm}
    \begin{equation*}
        \Xi_N^\mathcal{T} = \left\{ \xi_0  \; \middle| \; \exists u_{0, \dots, t}, \; \exists t \in \{0, \dots, N-1\}, \xi_t \in \mathcal{T} \right\}.
    \end{equation*}
\end{definition}

The definition of N-IOBRS closely resembles the backward reachable sets within the state space framework, with the key distinction that the state \( x \) is replaced by the extended state \( \xi \) to provide an implicit representation. Additionally, the index \( k \) indicates that the system's trajectory can reach the target set in fewer than \( N \) steps. \vspace{-0.1cm}
\subsection{Set-Theoretic MPC}
To elucidate the proposed data-driven method, we summarize the Set-Theoretic MPC (ST-MPC) approach in state-space framework, commonly known as dual-mode MPC \cite{angeli2008ellipsoidal}, with the assumption of one-step prediction and incorporating the equilibrium point as the first element of the nested backward reachable sets. Given the system \eqref{eq:LTI-system} and satisfying input-state constraints \eqref{eq:polytopic-constraints}, a receding-horizon controller capable of stabilizing the origin in finite time can be designed through the following offline and online steps:
\begin{enumerate}
    \item Offline - Define $\mathcal{T}^0 := 0 \in \mathbb{R}^n$ as the equilibrium point of system \eqref{eq:LTI-system}, then recursively compute a sequence of $n^*$ nested backward reachable sets, $\left\{\mathcal{T}^l\right\}^{n^*}_{l=1},$ where $\mathcal{T}^l\! =\! \{x \in \mathcal{X}\!:\! \exists u \in \mathcal{U}, Ax_t + Bu \in \mathcal{T}^{l-1}\}$.
    \item Online - Find  $l :=  \underset{l}\min \{l: x_t \in \mathcal{T}^l \}$ and solve \\$u_t = \underset{u}\min \, J(x_t,u) \text{ s.t. } x = Ax_t + Bu \in \mathcal{T}^{l - 1}$, where $J(x_t,u)$ is a convex function.
\end{enumerate}

\begin{figure}[b]
    \centering
    \includegraphics[trim=0cm 0cm 0cm 0cm, 
    clip,width=0.6\linewidth]{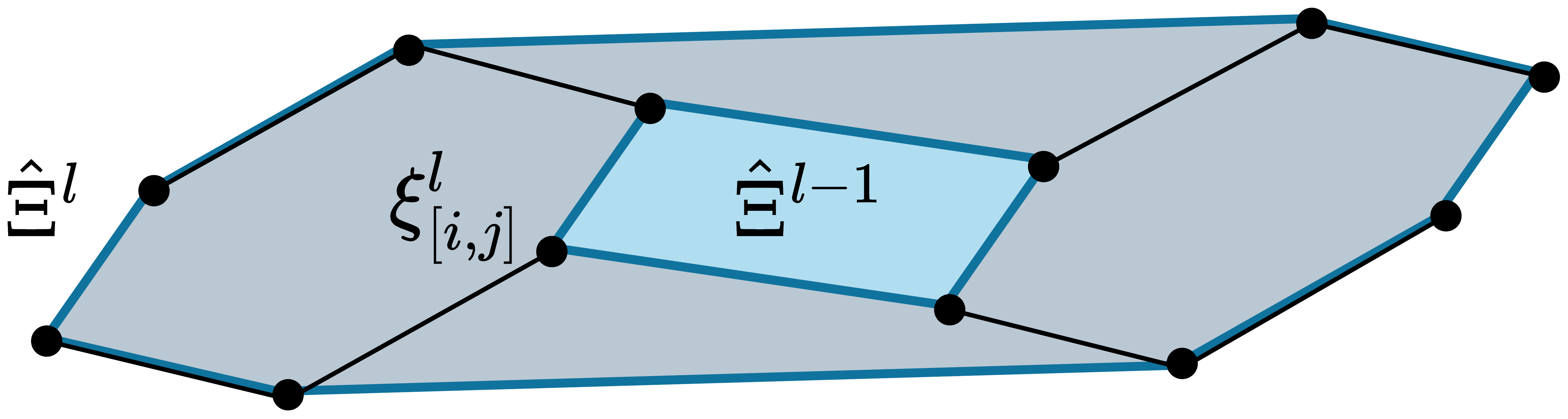}
    \caption{A visualization example of sample-based N-step input-output reachable sets for two nested sets: $\hat{\Xi}^{l-1}$ is the target set, $\hat{\Xi}^l$ is the corresponding N-IOBRS, $\xi_{[i,j]}^l$ is the $j^{th}$ element of $i^{th}$ sampled extended trajectory $\bar{\xi}_{i}^l$.}
    \label{fig:Set-Expansion_NIOBRS}
\end{figure}

\subsection{Problem of interest}

For the remainder of the paper, we will adopt the following assumptions to address Problem \ref{problem: ST-DDPC}, which are standard within the direct data-driven framework.

\begin{assumption}[Upper Bound on System's Lag] \label{Assumption: I} An upper bound on the system's lag is known $T_{\text{ini}}\geq \underline{l}$.
\end{assumption}

\begin{assumption}[Prediction Horizon Length] \label{Assumption: II}
The prediction horizon $N>2T_{\text{ini}}$.
\end{assumption}

\begin{assumption}[Persistent Excitation \cite{berberich2020robust}] \label{Assumption: III}
The stacked Hankel matrix \eqref{BST model} is PE of order $L = N+2T_{\text{ini}}$ in the sense of Definition \ref{Def: PE condition}.
\end{assumption}

\begin{assumption}[Equilibrium Point] \label{Assumption: IV}
An equilibrium point of the system (\ref{eq:LTI-system}) is the origin, and $(u^s, y^s)=(0, 0)$ belongs to admissible sets $(\mathcal{U},\mathcal{Y})$.
\end{assumption}

\begin{problem} \label{problem: ST-DDPC}
    Given an offline collected single input-output trajectory of system \eqref{eq:LTI-system} and fulfilling the assumptions (\ref{Assumption: I}-\ref{Assumption: IV}), design a direct data-driven predictive controller that ensures the input-output constraints \eqref{eq:polytopic-constraints} are always satisfied while the equilibrium point in the origin is reached in a finite number of steps.
\end{problem}

It is important to highlight that the proposed approach retains the advantages of DDPC, including multi-step prediction, the ability to handle input delays, implicit representation, and input-output measurements. Additionally, it benefits from all the properties of ST-MPC as outlined in \cite[Property 1, Sec II.B]{attar2023data} for the deterministic setting. We emphasize that although the proposed method is derived using behavioral system theory, it can also be reformulated to align with model-based approaches that utilize explicit multi-step input-output representations, such as Subspace Predictive Control (SPC) discussed in \cite{9654975}. In the following, we present a fully data-driven sampling approach for calculating the N-step input-output backward reachable set, as outlined in Section \ref{sec: N-IOBRS}, and then extend the ST-MPC framework to the input-output data-driven setup under consideration, as detailed in Section \ref{sec: ST-DDPC}.

\section{Sample-based N-step Input-output Backward Reachable Sets} \label{sec: N-IOBRS}
To approximate N-step Input-Output Reachable Sets (N-IOBRS), we build upon the concept of sample-based sets introduced in our recent study \cite{bajelani2024raw} within the input-output framework. Sample-based method has originally proposed in \cite{rosolia2017learning_2, wabersich2018linear} to expand the feasible set of MPC and model-based safety filters in the state-space framework. We underapproximate nested N-IOBRS in the sense of Definition \ref{def: N-IOBRS}. Consider the origin as the target set, $\xi^s = 0$, which is the equilibrium point of system \eqref{eq:LTI-system}. Also, consider ${\Xi}_N^{l=1}$ as a set of extended states that can be driven to $\xi^s = 0$ in at most $N$ steps. Since ${\Xi}_N^{l=1}$ is a convex set, any convex combination of sampled points from this set serves as its under-approximation. Generally, given \({\Xi}_N^{l-1}\) as a target set and \(\xi_{i,j}^l\) representing the \(j^{\text{th}}\) element of the \(i^{\text{th}}\) sampled extended trajectory \(\bar{\xi}_{i}^l\) belonging to the set \({\Xi}_N^{l}\), the under-approximation of \({\Xi}_N^{l}\), denoted as \(\hat{\Xi}_N^l\), is defined as follows:
\begin{equation}\label{eq:convexset-N-IOBRS}
\hat{\Xi}_N^l=\text{conv}(\{\bigcup_{i=1}^{N_i}\bigcup_{j=1}^{N_j} \xi_{i,j}^l, V({\Xi}_N^{l-1})\}),
\end{equation}
where $N_i$ and $N_j$ are the number of sampled extended trajectory and number of extended states in each sampled extended trajectory. Note that to under approximate N-IOBRS, $N_j = N$ and $\xi_{i,N}^l \in \Xi_N^{l-1}$. To under approximate the nested N-IOBRS, we use the data-driven safety filter formulation proposed in our recent work \cite{bajelani2024raw} to safely sample extended trajectories as follows:
\begin{subequations}\label{eq:DDSF}
\begin{align}
    &\underset{\substack{\alpha(t),\bar{u}(t),\bar{y}(t)}}{\min} \quad  \lVert \bar{u}_0(t) - u_{l}(t) \rVert_R^2 \label{eq:DDSF_cost} \\
    \text{s.t.} \quad & \begin{bmatrix}
        \bar{u}(t) \\
        \bar{y}(t)
    \end{bmatrix} = \begin{bmatrix}
        H_{L}(u^d) \\
        H_{L}(y^d)
    \end{bmatrix} \alpha(t), \label{eq:Hankel_DDSF} \\
    & \begin{bmatrix}
        \bar{u}_{[-T_\text{ini}, -1]}(t) \\
        \bar{y}_{[-T_\text{ini}, -1]}(t)
    \end{bmatrix} = \begin{bmatrix}
        u_{[t-T_\text{ini}, t-1]} \\
        y_{[t-T_\text{ini}, t-1]}
    \end{bmatrix}, \label{eq:initial_Condition_DDSF} \\
    & \begin{bmatrix}
        \bar{u}_{[N-T_\text{ini}, N-1]}(t) \\
        \bar{y}_{[N-T_\text{ini}, N-1]}(t)
    \end{bmatrix}  \in \Xi_N^{l-1},  \label{eq:Final} \\
    & \bar{u}_k(t) \in \mathcal{U}, \bar{y}_k(t) \in \mathcal{Y}, k \in \{0, \ldots, N-1\}.  \label{eq:I-O C_DDSF}
\end{align}
\end{subequations}
where $\alpha(t) \in \mathbb{R}^{N_0-L+1}$, $\bar{u}(t) = [\bar{u}_{-T_{\text{ini}}}^{\top},\cdots,\bar{u}_{N-1}^{\top}]^{\top}$, $\bar{y}(t) = [\bar{y}_{-T_{\text{ini}}}  ^{\top},\cdots,\bar{y}_{N-1}^{\top}]^{\top}$ are decision variables. Additionally, \(\bar{u}_k(t)\) and \(\bar{y}_k(t)\) indicate the \(k^{\text{th}}\) element of \(\bar{u}(t)\) and \(\bar{y}(t)\), respectively. The terms \(\bar{u}_{[-T_\text{ini}, -1]}(t)\) and \(\bar{y}_{[-T_\text{ini}, -1]}(t)\) represent \([\bar{u}_{-T_{\text{ini}}}^{\top}, \ldots, \bar{u}_{-1}^{\top}]^{\top}\) and \([\bar{y}_{-T_{\text{ini}}}^{\top}, \ldots, \bar{y}_{-1}^{\top}]^{\top}\) at time \(t\). The past input-output measurements with the length of $T_{\text{ini}}$ are denoted by ${u}_{[-T_\text{ini}, -1]} = [ {u}_{-T_{\text{ini}}}^{\top},\cdots, {u}_{-1}^{\top}]^{\top}$, $ {y}_{[-T_\text{ini}, -1]} = [ {y}_{-T_{\text{ini}}}^{\top},\cdots, {y}_{-1}^{\top}]^{\top}$ in \eqref{eq:initial_Condition_DDSF}, which implicitly characterizes the underlying state. Furthermore, \( u_l \) represents a random or learning input, \( \bar{u}_0(t) \) denotes the safe input applied to the system \eqref{eq:LTI-system}, and \eqref{eq:Hankel_DDSF} represents the implicit data-driven representation. Equation \eqref{eq:Final} imposes the terminal constraint on the last \( T_{\text{ini}} \) elements of \( [ \bar{u}(t), \bar{y}(t) ] \) to be in the set \( \Xi^{l-1}_N \), ensuring the safety of the filter. The input-output constraints are also defined by \eqref{eq:I-O C_DDSF}. The trajectory \( [ \bar{u}(t), \bar{y}(t) ] \) represents the backup input-output trajectory obtained from solving problem \eqref{eq:DDSF}, and it must be reshaped into an extended trajectory using Definition \ref{Extended Backup Trajectory}, to approximate \( \hat{\Xi}^l \). The proposed algorithm to calculate $n^*$ nested N-IOBRS is defined in Algorithm \ref{Algorithm:N-IOBRS}, and Fig. \ref{fig:Set-Expansion_NIOBRS} visualize an example of sampled nest N-IOBRS.
\begin{algorithm}
\begin{algorithmic}[1]
\caption{Sample-based nested N-IOBRS}\label{Algorithm:N-IOBRS}
\State \textbf{Input:} $u^d$, $y^d$, $N$, $N_i$, $T_{\text{ini}}$, $n^*$, and $\Xi^{0}$.
\State \textbf{Output:} $\Xi_N^{l=1:n^*}$.
\State Initialization $l=1$, $\xi(t=0) \in \Xi^{0}$
\For{$l = 1$ to $n^*$}
\State Solve problem \eqref{eq:DDSF} for ${N_i}$ steps.
\State Expand the safe set using \eqref{eq:convexset-N-IOBRS}.
\EndFor
\end{algorithmic}
\end{algorithm}
\begin{definition}[Extended Backup Trajectory]\label{Extended Backup Trajectory}
    The safety filter's backup trajectory is defined as the input-output trajectory provided by the solution of problem \eqref{eq:DDSF}. At time $t$, the extended backup trajectory is defined by \eqref{eq: Extended Trajectory} using $\left\{\bar{u}_k(t), \bar{y}_k(t)\right\}_{k=-T_{\text{ini}}}^{N-1}$.
\end{definition}
\begin{remark} [Offline Sampling of Nested N-IOBRS] It is also possible to compute the nested N-IOBRS offline using the implicit data-driven representation in \eqref{BST model} for one-step-ahead predictions. For details, refer to the offline set expansion algorithm in \cite{bajelani2024raw}.
\end{remark}
\begin{remark} [Nested property]
    Each iteration in Algorithm \ref{Algorithm:N-IOBRS} returns an under-approximated N-IOBRS, assuming the target set is the previous N-IOBRS. Based on \eqref{eq:convexset-N-IOBRS}, the resulting sets are nested, as each iteration accounts for the convex hull of the previous set's vertices and the new backup trajectories: \( \Xi^0 \subseteq \hat{\Xi}^1 \subseteq \dots \subseteq \hat{\Xi}^{n^{*}} \), if \( {\Xi}^0 \) is non-empty.
\end{remark}
    
\section{Set-Theoretic Data-Driven Predictive Control} \label{sec: ST-DDPC}
In this section, set-theoretic DDPC, inspired by \cite{8795639,attar2023data,angeli2008ellipsoidal}, is introduced and built upon the sampled N-IOBRS in the last section. Assuming single input-output trajectory of system \eqref{eq:LTI-system} and input-output constraints \eqref{eq:polytopic-constraints}, fulfilling assumptions (\ref{Assumption: I}-\ref{Assumption: IV}), ST-DDPC is defined as follows:
\begin{subequations}\label{eq:STDDPC}
\begin{align}
    &\underset{\substack{\alpha(t),\bar{u}(t),\bar{y}(t)}}{\min} \quad  \sum_{k=0}^{N-1} \|\bar{y}_k\|_{Q_y}^2 + \|\bar{u}_k\|_{Q_u}^2 \label{eq:STDDPC_cost} \\
    \text{s.t.} \quad & \begin{bmatrix}
        \bar{u}(t) \\
        \bar{y}(t)
    \end{bmatrix} = \begin{bmatrix}
        H_{L}(u^d) \\
        H_{L}(y^d)
    \end{bmatrix} \alpha(t), \label{eq:Hankel} \\
    & \begin{bmatrix}
        \bar{u}_{[-T_\text{ini}, -1]}(t) \\
        \bar{y}_{[-T_\text{ini}, -1]}(t)
    \end{bmatrix} = \begin{bmatrix}
        u_{[t-T_\text{ini}, t-1]} \\
        y_{[t-T_\text{ini}, t-1]}
    \end{bmatrix} \in \Xi_N^l, \label{eq:initial_Condition} \\
    & \xi_{[1:w-1]}  \in \Xi_N^{l}, \quad \xi_{[w:N-1]} \in \Xi_N^{l-1},\label{const:w}
\end{align}
\end{subequations}
where $\xi_{[1:w-1]}$ and $\xi_{[w:N-1]}$ is defined as follows:
\begin{equation*}
    \xi_{[1:w-1]} = \begin{bmatrix}
        \bar{u}_{[k_1-T_\text{ini}, k_1-1]} \\
        \bar{y}_{[k_1-T_\text{ini}, k_1-1]}
    \end{bmatrix}, \quad k_1 \in \{1, \ldots, w-1\},
\end{equation*}
\begin{equation*}
    \xi_{[w:N-1]} = \begin{bmatrix}
        \bar{u}_{[k_2-T_\text{ini}, k_2-1]} \\
        \bar{y}_{[k_2-T_\text{ini}, k_2-1]}
    \end{bmatrix}, \quad k_2 \in \{w, \ldots, N-1\}.
\end{equation*}

Furthermore, $w$ is the length of a sliding window, which is updated in each time step using the Algorithm \ref{algorithm:set_theoretic_MPC}. Note that the sliding window in \eqref{const:w} forces the last $w$ element of the extended prediction trajectory to be within the next set. For points farthest from \( \Xi_N^{l-1} \), \( w \) must be \( N-1 \); that is, reaching the next set requires at most \( N \) steps, see Definition \ref{def: N-IOBRS}. Accordingly, Algorithm $w$ starts from $N-1$ and decreases by one $w \gets w-1$ at each time step. Shrinking the window's length guarantees even the farthest point in $\Xi_N^{l-1}$ enter the next set at most after $N$ steps. Since it would also be possible to enter the next set less than $N$ step based on Definition \ref{def: N-IOBRS}, $w$ must be also reset to $N-1$ once the extended state derived to $\Xi_N^{l-1}$.
\begin{algorithm}
\caption{Set-Theoretic DDPC \label{algorithm:set_theoretic_MPC}}
\begin{algorithmic}[1]
\State Initialization: $u^d$, $y^d$, $\Xi^{j=1,\cdots,n^*}$, $N$, $w=N-1, \xi(0)$.
\State Find the smallest $\Xi^j$ that contains $\xi(0)$ by solving \Statex $l := \displaystyle \min_{l\in \{0,\ldots,n^*\}} \{l: \xi(0) \in \Xi^l\}$
\For{$t = 1$ to $lN$}
\State Solve ST-DDPC problem \eqref{eq:STDDPC} to calculate $u_{0}(t)$.
\State Apply $u_{0}(t)$ to the system \eqref{eq:LTI-system}.
\State Find the smallest $\Xi^l$ that contains $\xi(t)$ by solving \Statex $l := \displaystyle \min_{l\in \{0,\ldots,n^*\}} \{l: \xi(t) \in \Xi^l\}$
\State Update the sliding window \( w \gets w - 1 \)
\Statex
If { \( w = 0 \) or \( \xi(t+1) \in  \Xi^{j-1} \)} then \( w \gets N-1 \).
\EndFor
\end{algorithmic}
\end{algorithm}
\begin{theorem} [Recursive Feasibility and Convergence] Let assumptions (\ref{Assumption: I}-\ref{Assumption: IV}) hold. If the problem \eqref{eq:STDDPC} is feasible at \( t = t_0 \), then it remains feasible for all \( t > t_0 \). Additionally, the system can reach the origin in finite time.
\end{theorem}
\textbf{Proof.} Suppose the initial condition belongs to set $\Xi_N^{l}$. Since, by the definition of N-IOBRS, the solver is able to guide the system to the next set at most in $N$ steps to the target set $\Xi_N^{l-1}$, then the problem is feasible at least for $N$ steps. By shifting the argument $\Xi_N^{l} \to \Xi_N^{l-1}$ as the initial set and $\Xi_N^{l-1} \to \Xi_N^{l-2}$ as the target set, it is possible to conclude the problem stays feasible for all future time steps by relying on induction. This also shows that since the problem is recursively feasible, for any initial condition in $\Xi_N^{l}$, it takes at most $N$ steps to enter the next set. For any initial condition in \( \Xi_N^{l} \), the maximum convergence time is \( lN \) based on the definition of nested N-IOBRS. \hfill$\square$

Note that the proof of recursive feasibility and convergence results in the proof of the constraint satisfaction. Since, by construction, if the problem \eqref{eq:STDDPC} is feasible at \( t=0 \), then input-output constraints are respected for infinite time.

\section{Numerical Results and Discussion} \label{sec: results}

To highlight the effectiveness of the proposed method, we make a comparison between ST-DDPC and DDPC, proposed in \cite{9654975}, using the following unstable system:
\begin{equation}
\left[\begin{array}{l|l}
A & B \\
\hline C & D
\end{array}\right]=\left[\begin{array}{cc|c}
1 & 1 & 0\\
0 & 2 & 1\\
\hline 1 & 0 & 0\\
\end{array}\right].
\end{equation}

\begin{figure}[t]
  \centering
  \includegraphics[width=1\linewidth]{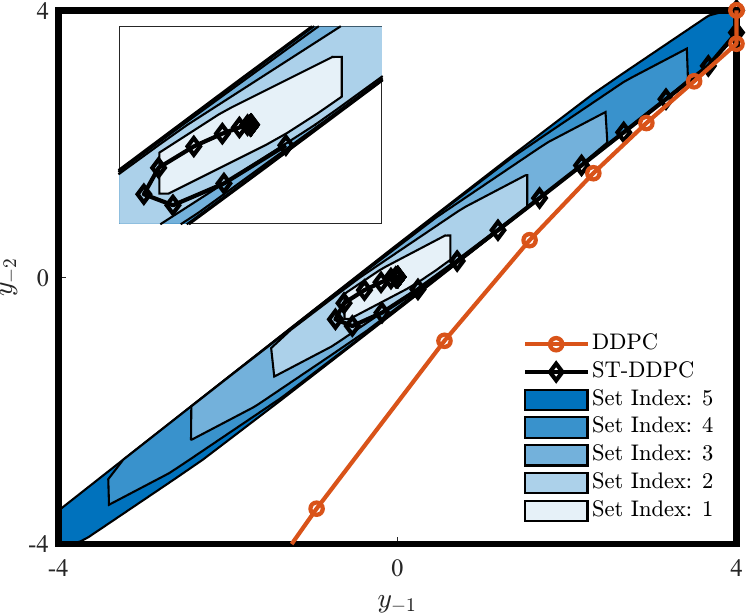}
\caption{Projection of N-step Input-Output Backward Reachable Sets (N-IOBRS) onto the output space, \(Proj_{[y_{-1},y_{-2}]}(\Xi^{1}), \ldots, Proj_{[y_{-1},y_{-2}]}(\Xi^{5})\), alongside the realized extended trajectories of DDPC and ST-DDPC.}
  \label{fig: Backwards_DDPC_ST_DDPC}
\end{figure}

To ensure a fair comparison, identical weights are selected for both DDPC and ST-DDPC, with $Q_y = I_{2 \times 2}$ and $Q_u = 1$. The input and output constraints are defined as $\lvert u \rvert < 0.5$ and $\lvert y \rvert < 4$, respectively. The prediction horizon is defined as $N = N_p + T_\text{ini}$, where $N_p = 4$ and $T_\text{ini} = 2$. The matrices $H_L(u^d)$ and $H_L(y^d)$ are derived from a single trajectory of the open-loop system, which satisfies the PE condition stated in Assumption \ref{Assumption: II}. By applying the sample-based method described in Section \ref{sec: N-IOBRS}, the N-step input-output backward reachable sets are computed, and their projections onto the subspaces of past outputs are depicted in Fig. \ref{fig: Backwards_DDPC_ST_DDPC}. Note that the actual N-IOBRS exists in \(\mathbb{R}^{4}\); for \(T_\text{ini}=2\). Since the extended state is defined as the two past input-output data \(\xi = [u_{[-1,-2]}^{\top},y_{[-1,-2]}^{\top}]^{\top}\).

\begin{figure}[t]
  \centering
  \includegraphics[width=1\linewidth]{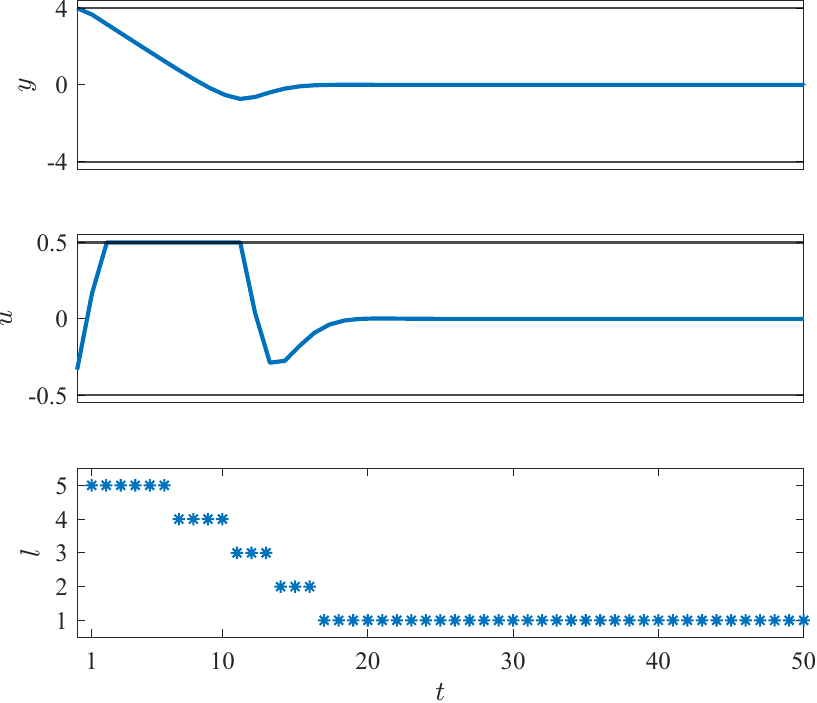}   
    \caption{Input-output trajectory of the proposed method, ST-DDPC, along with the corresponding realized set index.}
  \label{fig: ST_DDPC}
\end{figure}

\begin{figure}[t]
  \centering
  \includegraphics[trim=0cm 0cm 0cm 0cm, 
    clip,width=0.95\linewidth]{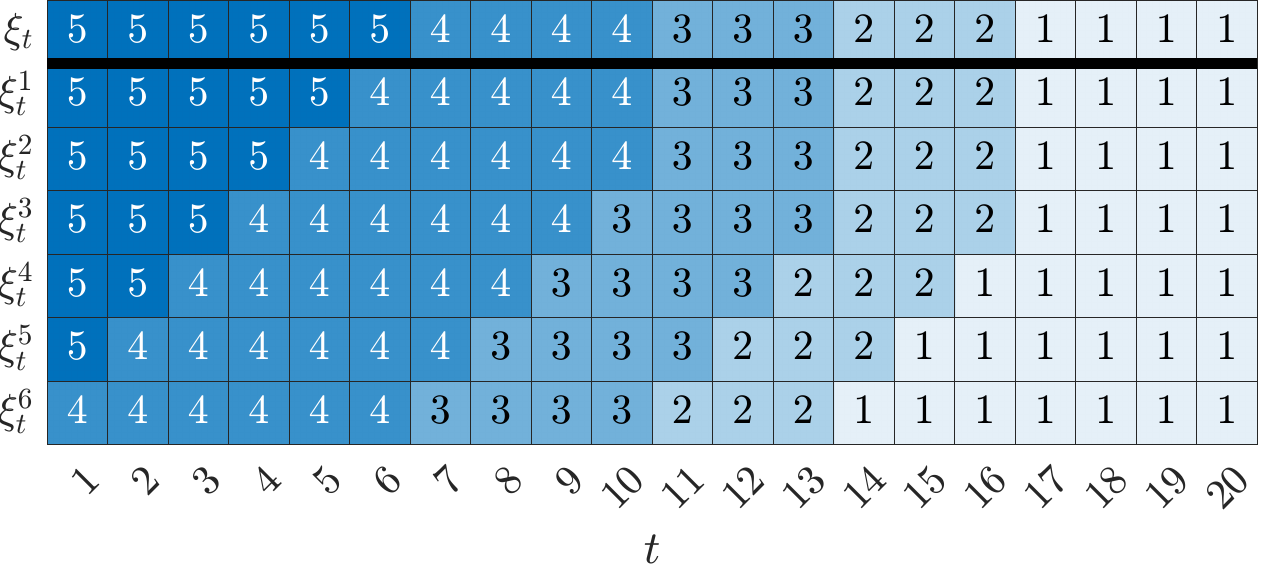}
\caption{Sliding window over the prediction horizon: the first row represents the realized extended state set, while the remaining rows depict the set constraints on the extended predicted trajectory at time $t$, as determined by Algorithm \ref{algorithm:set_theoretic_MPC}.}
  \label{fig: Sliding_Window}
\end{figure}

Assuming $x_0 = [4, 0]^\top$, or equivalently $y_{ini} = [4, 4]^\top$ and $u_{ini} = [0, 0]^\top$, the realized trajectories of the system under both DDPC and ST-DDPC strategy are implemented by CasADi and MPT3 toolboxes \cite{Andersson2018,MPT3} and shown in Fig. \ref{fig: Backwards_DDPC_ST_DDPC}. Since DDPC cannot stabilize the system, the simulation is shown only for 8 time steps. However, ST-DDPC successfully guides the system to $y_{ini} \approx [0, 0]^\top$ and $u_{ini} \approx [0, 0]^\top$, or equivalently $x_0 \approx [0, 0]^\top$, while respecting the input-output constraints and the N-IOBRS. Note that due to numerical error associated with the solver, it is hard to exactly achieve $x_0 = [0, 0]^\top$. It is recommended to use the data-driven output feedback controllers proposed in \cite{qin2024data,alsalti2023notes} to define $\hat{\Xi}^1$ for obtaining a robust numerical solution; in this case, the set $\hat{\Xi}^1$ will exhibit finite-time convergence. The input-output trajectories, input-output constraints, and the realized set index are illustrated in Fig. \ref{fig: ST_DDPC}. Additionally, the sliding window length, defined by the parameter \(w\), is shown in Fig. \ref{fig: Sliding_Window}. Please visit the \href{https://www.youtube.com/watch?v=wDQZ7UfKcZE}{\emph{link}} \footnote{\href{https://www.youtube.com/watch?v=wDQZ7UfKcZE}{https://www.youtube.com/watch?v=wDQZ7UfKcZE}}
  for a video of the simulation, including the sample-based N-IOBRS.


It is important to recognize that while ST-DDPC achieves stabilization in a finite number of steps regardless of the parameter values for \(Q_y\) and \(Q_u\), DDPC can stabilize the system only for a proper choice of these parameters. Although DDPC with a long prediction horizon and a well-tuned objective function may achieve stabilization without terminal constraints, this approach tends to be computationally demanding, as the size of the Hankel matrix increases with the prediction horizon. Moreover, tuning \(Q_y\) and \(Q_u\) to ensure stability is not trivial in the absence of terminal ingredients.

\section{Conclusion} \label{sec: Conclusion}

In this paper, a set-theoretic approach is proposed to ensure the recursive feasibility and finite-time convergence of DDPC without the need for an explicit model of the system and explicit state estimation. The entire process, from sampling nested backward reachable sets to designing set-theoretic predictive control, is purely data-driven and ensures that input-output constraints are satisfied. This work also demonstrates how safety filters can be integrated into the predictive controller design process. A key direction for future research is to address the impact of measurement noise on Hankel matrices and initial conditions, as it directly influences prediction accuracy. In this case, one approach is to robustify the proposed method by tightening the backward reachable sets and introducing regularization into the objective function.

\bibliographystyle{IEEEtran}
\balance
\bibliography{references}

\end{document}